\documentstyle[aps,multicol,epsf]{revtex}

\def\L{{\cal L}}
\def\H{{\cal H}}
\def\eps{\epsilon}
\def\ep0{\epsilon^0}

\def\x{\epsilon_R}
\def\y{\epsilon_I }
\def\z{a}
\def\z1{b}

\def\be{\begin{equation}}
\def\ee{\end{equation}}

\begin{document}
\title{ Winding Numbers, Complex Currents,  and  Non-Hermitian Localization  }
\author{Nadav M.  Shnerb and  David  R. Nelson}
\address{ Lyman Laboratory  of Physics, Harvard  University,
Cambridge, MA 02138}
\date{\today}
\maketitle

\begin{abstract}
The nature of extended states in disordered tight binding models 
with a constant imaginary vector potential is explored. Such models, 
relevant to vortex physics in superconductors and to population biology, 
exhibit a delocalization transition and a band of extended states even for 
a one dimensional ring. Using  an   analysis of  eigenvalue trajectories
 in the complex plane, we 
demonstrate that each delocalized state is characterized by an (integer) 
winding number, and evaluate the 
associated  complex current. 
Winding  numbers in higher dimensions are also discussed.
\pacs{PACS: 72.15.Rn, 73.20.Jc,  74.60.Ge, 05.70.Ln }
\end{abstract}

\begin{multicols}{2}

There is  a growing interest in the spectra  of  
random non-Hermitian   matrices \cite{general}.
Considerable attention has focused on a particularly simple class 
of tight binding Anderson models 
with a constant imaginary vector potential, inspired by the physics 
of vortex matter \cite{hat}. 
These models exhibit a sharp delocalization transition 
even in one and two dimensions. Similar operators, represented by large real, 
asymmetric sparse matrices, arise in theories of population biology
in random media with convection \cite{NS}, and in many other contexts 
\cite{efetov}. Delocalized   eigenmodes arise in response
to a sufficiently large asymmetry parameter, accompanied by 
eigenvalues which escape in conjugate pairs from the real axis into the 
complex plane  \cite{hat}.

Although   the initial work relied  heavily on  numerical investigations 
\cite{hat}, there has been considerable recent analytic progress, 
particularly in one dimension. Brouwer et. al. calculate 
explicitly finite size effects and the bubble of complex eigenvalues
 representing extended states in the center  of the band in the limit of 
weak disorder \cite{brouwer}.  
Brezin and Zee \cite{brezin} have obtained exact results 
for the density of states for the special case of Lorentzian site randomness
embodied  in the Lloyd model \cite{Lloyd}.
 Goldsheid and Khoruzhenko  discuss this  model and  
 show more 
generally that the eigenvalues are distributed along curves in the 
complex plane, providing  analytic formulas relating the spectrum to the 
properties of a  
reference Hamiltonian with no asymmetry \cite{russim}. The nontrivial 
behavior which results  for {\it large} asymmetry parameters 
in two and three dimensions has been studied  analytically via 
renormalization group calculations and a mapping onto Burgers equation 
\cite{NS}.

In this paper we study the eigenfunctions and complex currents associated with 
the band of extended states in one dimension. Unlike delocalized states in 
Hermitian disordered systems (where the eigenfunctions can  always be
chosen to be  real), we show that these complex eigenfunctions 
are characterized by a conserved  winding number $n$ even when the disorder is 
 strong. Such topological quantum numbers 
can be used to label the eigenvalue spectrum $\epsilon_n(g)$,
 where $g$ is the
asymmetry parameter, i.e., the imaginary vector potential. 
A study of the eigenvalue trajectories as a function of 
$g$ then leads to complex currents 
$J_n = -i {\partial \epsilon_n \over \partial g}$, which determine  
the average tilt $dr/dz$ of the vortex trajectory $r(z)$ in a superconductor
with columnar pins in the presence of a perpendicular  magnetic 
field proportional to 
$g$ \cite{hat}. Assuming for simplicity a periodic box of size $L_z$ in the 
column direction, one has, for example, 
\be
<{d r(z) \over dz}> = - i  {\sum_n  J_n e^{-\epsilon_n L_z /T}
\over \sum_n e^{-\epsilon_n L_z /T}   },
\ee
where the brackets denote a thermal average at temperature $T$. 
We shall also  argue that winding numbers can  be used to label extended 
states for {\it large} asymmetry parameters in two or more  dimensions.

The   one dimensional non-Hermitian  
tight binding  model ``Hamiltonian'' in a basis of 
$N$  sites localized at positions  $\{ r_j  \} $  reads,
\begin{eqnarray}\label{1} 
\H_{i,j} = -{1 \over 2}  
w_j   \left( e^{h_j }  \delta_{i+1,j} + 
e^{-h_j  }  \delta_{i-1,j}    \right) 
 -  V_j \delta_{ij},
\end{eqnarray}
where we assume periodic boundary conditions 
and the  $\{ h_j \}$ are  real asymmetry parameters. 
When exponentiated, this operator describes the transfer matrix 
for a  flux line with columnar defects in a cylindrical
shell \cite{hat}. The Liouville operator describing 
population growth on a lattice  in an inhomogeneous environment with 
convection is given by $\L = - \H$ \cite{NS}. 
The zero mean  random potential $V_j$ is chosen independently for each site,
and arises from the variations of columnar pin 
diameters (vortex matter) or inhomogeneous growth rates (population biology).
As discussed in \cite{hat}, the randomness in the off-diagonal hopping is due
to the irregular  
spacing between columnar defects in superconductors. If $a_j$ is the 
spacing between nearest neighbor columns $j$ and $j+1$, then the
 $w_j = w_0
 \exp(-\gamma  a_j)$ and $h_j = g a_j/a$, where $\gamma$ and $w_0 > 0$
are constants, $a$ is the average lattice constant, and $g$ is proportional 
to the field component which tries to 
tip vortices away from the columns. In models of population growth,
$w_j$ describes diffusion between inhomogeneously distributed 
population centers and $h_j$ represents a fluctuating convection 
velocity with average value $g$ and second moment $\sigma$. 
We impose  periodic boundary conditions in the space-like direction 
i.e.,  the $N+1$-st site is identified  with the first site of the 
sample. 
     
Upon carrying out a similarity transformation on the 
Liouville operator, 
$\L'  =  S \L S^{-1}$, where 

\be\label{3}
S_{i,j} = \delta_{i,j} \exp[\sum_{k=1}^{j-1} h_k   - {(j-1) \over N}
\sum_{k=1}^N h_k ] ,
\ee
one finds that 
the spectrum will be the same as that of a Liouvillian  with a 
{\it uniform} asymmetry parameter, namely $\L' = - \H'$ with 

\be \label{4}
\L'(g)  =  \left( 
\begin{array}{ccccc}
V_1 & {w_1 \over 2} e^{g}  &0    &..&{w_N \over 2}   e^{-  g }  \\
{w_1 \over 2} e^{-g}  & V_2 &{w_2 \over 2}e^{g}   &..&0 \\
0& {w_2 \over 2} e^{-g}  &V_3&..&0 \\
:&:&:&:&: \\
{w_N \over 2}  e^{ g }    &0&0&.. &V_N \\
\end{array}
\right). 
\ee  
where ${g} = {1 \over N} \sum_{j=1}^N h_j $. Eqs. (\ref{3} - \ref{4})
generalizes the gauge transformation used to describe 
{\it localized} states in, e.g., Ref. \cite{hat} and \cite{NS}. 
The sample to sample fluctuations of ${g}$ about its  
 mean fall off  like 
$\sigma/\sqrt{N}$ \cite{note}.  As $N$ grows one may thus  replace 
the fluctuating quantity $h_i$ by a disorder independent average value 
 $g$, as we shall 
 do in the rest of this 
paper. This result implies 
{\it universality} in the response to random convection
in 1D growth models - nonuniform convection velocities may be mapped 
into a uniform average velocity via the  transformation
(\ref{3}), 
provided one applies this similarity transformation
to the eigenfunctions as well.

For a 1D ring with random parameters $\{V_i\} , \{w_i\}$ and    $g = 0$,
all the eigenfunctions of (\ref{4})  are real and localized, 
and its eigenvalues are real and discrete \cite{loc}. We assume a  large but 
finite chain, such that although the spectrum is discrete,  the 
length of the chain is much larger than the maximal localization length
of an eigenmode.

In a typical symmetric  one-dimensional  disordered system
with ${g} = 0 $, the localization 
length $\xi$ is largesy at the center of the band and smaller at the tails.
The criterion  for delocalization of the asymmetric system is $\kappa a < 
g$, where $\kappa \equiv 1/\xi$ \cite{CDW}.  
As a result,  pairs of complex energies representing  delocalized states
first appear as a ``bubble'' at the center of the band, which  then spreads
into the band 
tails.  
To study the complex currents associated with delocalized states 
we  follow Refs. [5-7] and exploit the  
relation between the complex spectrum of the  asymmetric problem
 with $g \neq 0$  and the real eigenvalues of a 
``background'' localized problem with  $g=0$, and  
same realization of disorder.
\begin{minipage}[t]{3.2in}
\epsfxsize=3.2in
\epsfbox{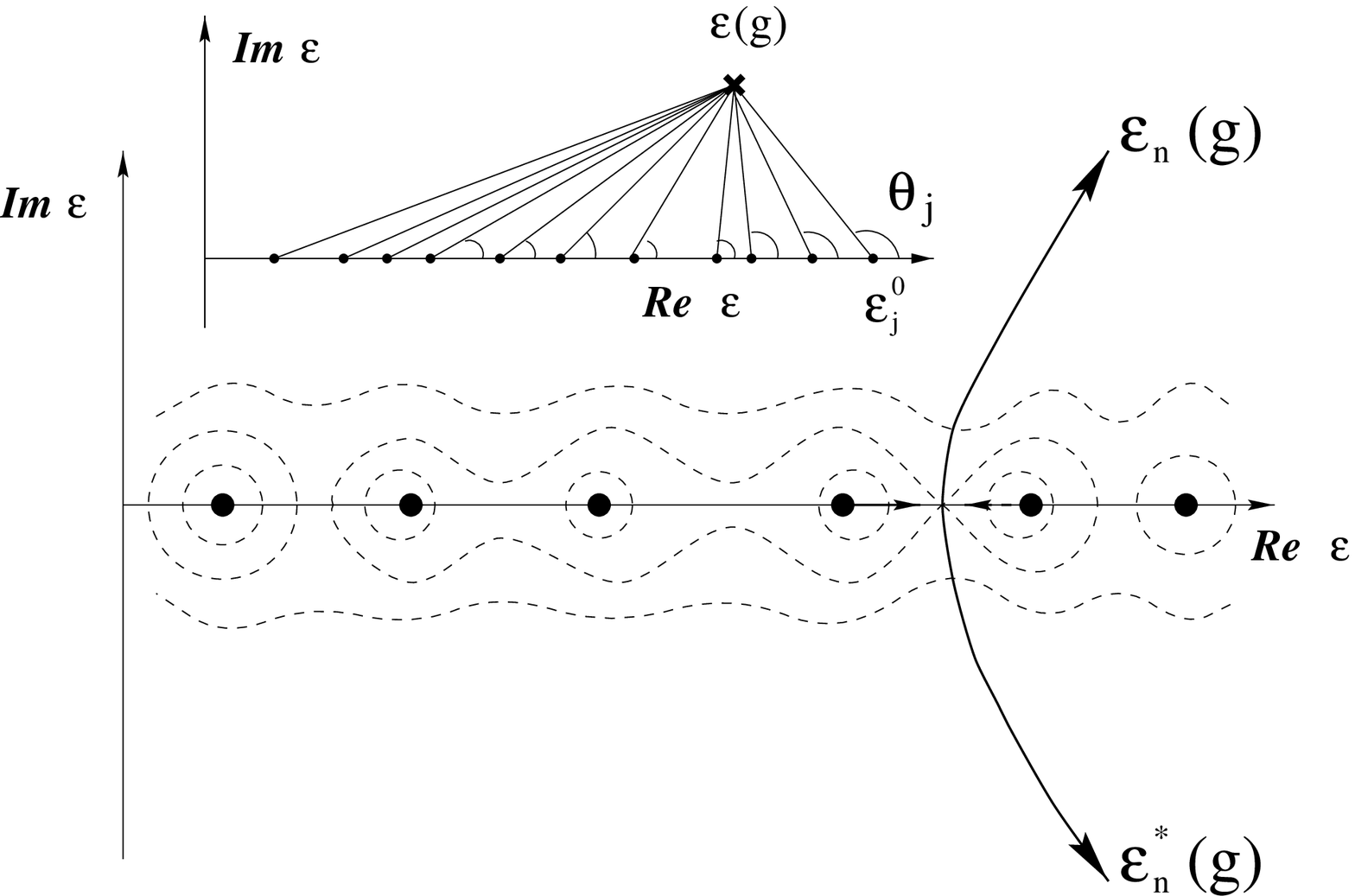}
\begin{small}
FIG.\ 1.  
Eigenvalue trajectories in the 
complex plane. Inset: Angles entering Eq.  (6) for 
a given $\epsilon(g)$.  
\end{small}
\vspace{0.2in}
\end{minipage}
      
The condition for a complex number $ \epsilon  = \x + i \y $ to be 
an eigenvalue of the matrix $\L'(g)$ is $Det[\eps-\L'(g)] = 0$. Eq.  
(\ref{4}) implies that $\eps$ is an eigenvalue only if 
\cite{brouwer,thouless} 
\begin{eqnarray} \label{thou}
Det[\eps-\L'(g=0)] &=& \prod_{i=1}^N (\eps - \ep0_i) \nonumber \\
 &=&  
   2 [\cosh(g N) - 1]  \prod_{j=1}^N  \left( {-w_j \over 2} \right)   
\end{eqnarray}
where the  $\{\ep0_i\}$ 
are the (real) eigenvalues of the background matrix  $\L'(g=0)$. 
To extract winding numbers, we first observe 
that  the right hand side 
of (\ref{thou}) is real and positive for even $N$, while for
odd $N$ it is real and negative. As a result,   
each complex $\eps$ should satisfy (see inset to Fig. 1)
\begin{eqnarray} \label{arg}
\sum_i cot^{-1} ({\x - \ep0_i \over \y}) = p \pi,
\end{eqnarray}
with  $p = 2n$ for even $N$, $p=2n+1$ for odd $N$, where $n$ is an integer,  
and the function 
$cot^{-1}(x)$  varies  from $\pi$ to $0$ as 
$x$ goes between $-\infty$ and $\infty$.

As $\y \to 0$,  each term in Eq. (\ref{arg})   
gives $\pi$ for every  eigenvalue $\epsilon_i^0$ 
 to the right of $\x$, and zero 
for each eigenvalue to the left. 
To satisfy (\ref{arg}) for even (odd) $N$ and a  given $n$, 
the eigenvalue must leave the real axis and enter  the complex plane 
at the gap between the 2n-th and the 2n+1-th (2n+1-th and the 2n-th)
 eigenvalues  of the 
$g=0$ ``background'' system.   We call $n$ the {\it index} of the 
trajectory $\x(g) + i \y(g) $ in the complex plane.

For odd  $N$, we see immediately from (\ref{arg}) that the rightmost 
eigenvalue (with  $p=n=0$) must remain real, consistent 
with Perron-Frobenius theorem  \cite{peron}. 
The corresponding nodeless eigenfunction corresponds to 
the ground state of $\H' = -\L'$. For $N$ even, particle-hole symmetry 
implies that both the rightmost and leftmost eigenvalues are always real. 
More generally, for a fixed value of $n$, the set of all 
$[\x(g), \y(g)]$ satisfying Eq. (\ref{arg}) 
defines a curve in the complex plane, as 
illustrated in Fig. 1. Henceforth, we assume $N$ even for simplicity.  

A more  complete description of the eigenvalue trajectory
results from 
taking the logarithm of the modulus  of Eq. (\ref{thou}).
In  the limit $N g >>1$ one finds
a second  constraint on $\eps (g) $, namely, [5-7]
\be \label{loc}
|g| = -\ln({w \over 2}) + {1 \over N} \sum_i \ln (|\eps - \ep0_i|) 
\ee
where $|\eps - \ep0_i|  = \sqrt{(\x-\ep0_i)^2 + \y^2}$ and $w = 2   
[\prod_{j=1}^N {w_j \over 2} ]^{1 \over N}$. 

This constraint is described graphically in Fig. 1, which shows schematically 
the level curves defined by (7) near the band center for three values of $g$. 
These are lines 
of constant potential for an equivalent 2d electrostatic problem with 
charges at the positions of the localized point spectrum for $g=0$. 
When $g$ is small, the constraint is solved by eigenvalue pairs on the 
real axis in the gaps between 
neighboring $\epsilon^0_j$'s indexed by $p=2n$. 
The departure of the eigenvalues from their $g=0$ values is  initially
 exponentially small in the system size \cite{hat}. As $g$ increases, however, 
successive pairs of eigenvalues eventually merge at a saddle point 
in the potential contours  
and detach from the real axis at right angle. For a given real  energy 
$\epsilon_R$,  Thouless has defined an energy dependent inverse localization 
length $\kappa(\x)$ for the associated Hermitian problem \cite{thouless},
 \be
\kappa(\x)  = -\ln({w \over 2}) + {1 \over N} \sum_i \ln (|\x - \ep0_i|).
\ee
Upon comparing with Eq. (\ref{loc}), we see that $\y$ becomes nonzero 
whenever $|g| > \kappa(\x')$ where $(\x', 0)$ is the detachment point 
of the eigenvalue pair. 

As $g$ increases above $\kappa(\x')$, Eqs. (\ref{arg}) and (\ref{loc}) 
define a unique pair of 
complex eigenvalue trajectories $\epsilon_n(g)$ and $\epsilon_n^*(g)$
for every value of $n$. 
Upon passing to the limit $N \to \infty$, the spectrum $\{ \epsilon_j^0 \}$
for $g=0$ closes up, and is described by a density of states $\rho_0(\lambda)$.
Eq. (\ref{arg}) and (\ref{loc}) 
may then recombined into a single complex equation, namely

\be\label{8}
\int_{-\infty}^{\infty} d \lambda \; \rho_0(\lambda) \ln[\epsilon_n (g)  
- \lambda]
 = \ln \left( {w \over 2} e^{|g|} \right)  + i \pi \left( {2 n \over N} \right).
\ee
where $\ln[\epsilon (g)  - \lambda] = \ln[|\x +i \y   - \lambda|] + i cot^{-1}
[(\x-\lambda)/\y]$.
In the limit $N >> 1$ and for  a density of states function 
symmetric around the special detachment point with $\x' = 0$, 
there is a purely imaginary trajectory of the form $\epsilon(g) 
= i \y(g)$ with $n = N/2$. For fixed $g$, Eq. (\ref{8}) thus  
leads to an implicit formula 
for the ``height'' $\y^{max}$ of the bubble of complex eigenvalues in the 
center of the band, namely,
${1 \over 2} \int d \lambda \;  \rho(\lambda) \;   
\ln [\lambda^2 + (\y^{max})^2] =
|g|   + \ln({w \over 2}).$ This integral vanished, as expected [6],  
in the ``one way'' limit, $g \to \infty$ with $ {w \over 2} e^{|g|} = 1$.
For other detachment points, the eigenvalue trajectories 
curve to the left or right as required by the constraint (6) 
(See Fig. 1). 

The analysis above  
suggests that the imaginary parts of most   $\{ \epsilon_n(g) \}$
diverge as $|g| \to \infty$. 
It is then expected that  {\it all}
 eigenfunctions $\phi_n(j)$ are approximately 
plane waves, $\phi_n(j) \sim \exp(ik_n r_j /a)$, with free particle eigenvalue 
energies $\epsilon_n(g) = w \cos(k_n + ig)$ \cite{hat,NS}. For large 
$|\y(g)|$,  Eq. (\ref{8}) leads to
\be
\epsilon_n(g) \approx  {w \over 2} \exp(|g| + 2 i \pi n/ N).
\ee
Comparison with the free particle spectrum at large $|g|$ shows immediately
  that the index $n$ of the eigenvalue trajectory and the wavevector are 
related, $k_n = 2  \pi n / N$. These  wave eigenfunctions spiral around the 
origin in the complex plane as one moves along the 1d lattice of the tight 
binding model  sites, leading to a well defined winding number $n$. 
The winding number associated with the eigenvalue 
$ \epsilon^*_n(g)$ in the lower half plane is then $-n$. 

As $g$ decreases, the associated delocalized wave function $\phi_n(j;g) $
must {\it remain} nonzero at every site: If $\phi_n(j;g) $ were in fact 
exactly zero at some site $i$, it can be shown that the state is localized 
with  a real eigenvalue and eigenfunction by mapping 
all effects of the asymmetry onto the special  site $i$ via a
transformation like 
(\ref{3}). 
Thus, the  winding number must be {\it preserved} as $g$ 
decreases, i.e., the winding number is a topological invariant along an 
eigenvalue trajectory. The projection of such a 
delocalized eigenfunction is illustrated in the inset to  
Fig. 2.

As $N \to \infty$, we can replace the winding index $n$ by the continuous
variable $k_n = {2 \pi \over N} n$. Eq. (\ref{8}) then shows quite generally 
that the complex spectrum is a function only of the combination 
$g+ik$, i.e., $\epsilon_n(g) = \epsilon(g+ik)$. It then follows from the 
Cauchy-Riemann relations that all eigenvalue trajectories are at 
right angles to the lines of constant $g$, ${d \epsilon(g) \over dg} = -i 
{d \epsilon(g) \over dk}$. 
\begin{minipage}[t]{3.2in}
\epsfxsize=3.2in
\epsfbox{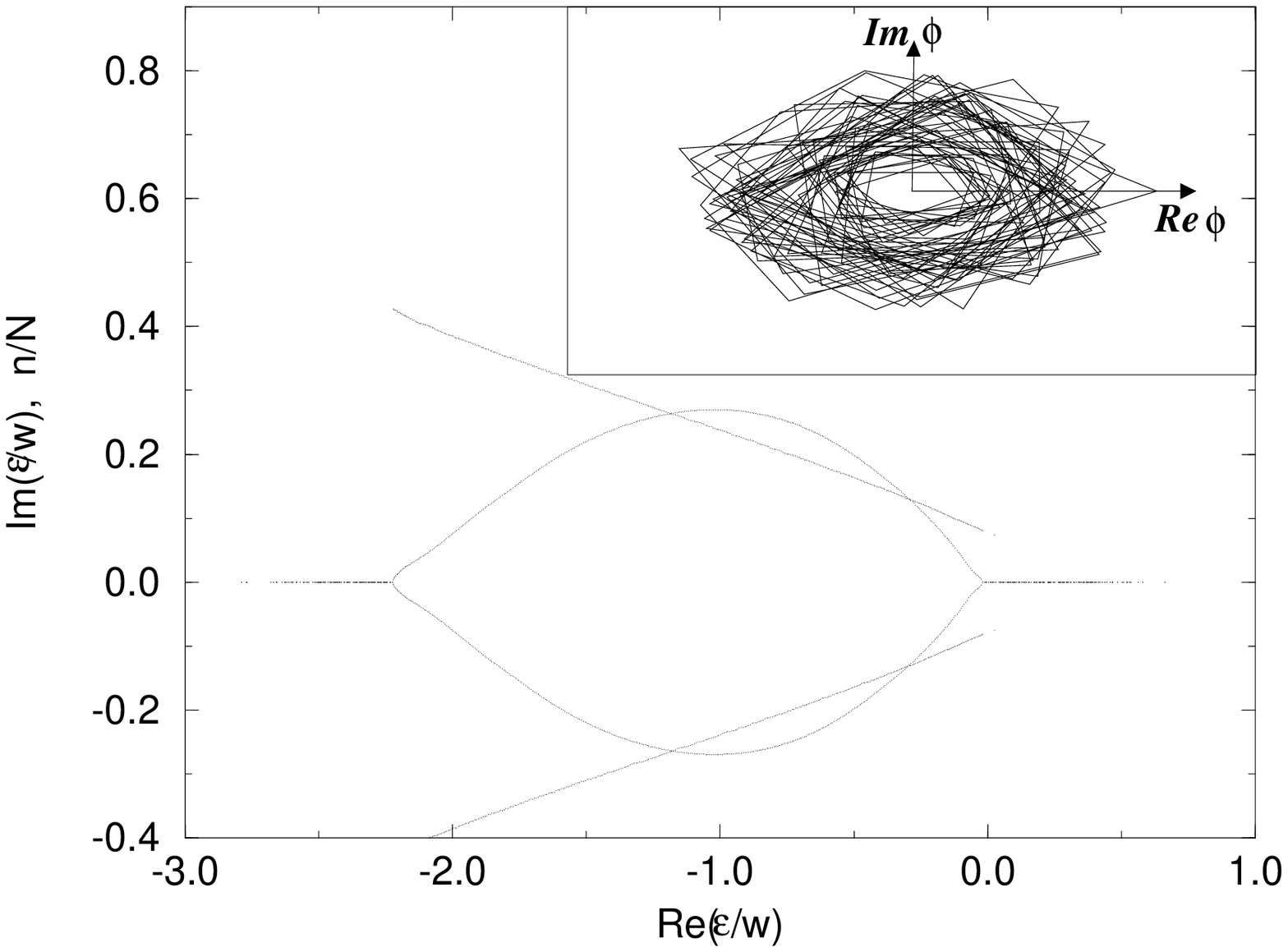}
\begin{small}
FIG.\ 2. 
Spectrum and  winding numbers 
for  $N = 1000$  with $V_j$ uniformly distributed 
in the interval $[-1,1]$,  and $g=0.4$. 
Inset: Projection onto the complex plane
of eigenfunction with $N = 200$, $g=1$,    and $n=100$.
\end{small}
\vspace{0.2in}
\end{minipage}

An explicit formula for the complex current 
results, moreover, from differentiating Eq. (\ref{8}) with respect to $g$,
\begin{eqnarray}\label{11}
J[\epsilon_n(g)] &=& -i {\partial  \epsilon_n(g) \over \partial g}
\nonumber \\ &=& 
\left[ i  \int d\lambda \; { \rho_0(\lambda)  \over \epsilon_n(g) - \lambda}  
\right]^{-1} \equiv -i G_0^{-1} [\epsilon_n(g)]
\end{eqnarray}

Evidently, the current associated with a particular complex eigenvalue 
$\epsilon_n(g)$ is determined by the Green's  function of the 
$g=0$ problem. As an application of our results, 
we note that the disorder averaged  Green's function for the Lloyd 
model, with Lorentzian site disorder,
$Pr(V_j) = {1 \over \pi} { \gamma \over (V_j)^2 + \gamma^2}$ is 
\cite{brezin,Lloyd} $<G_0(E)> = \left[(E+i\gamma) - t \right]^{-1/2}$,
and complex energy spectrum for extended states with ${\rm Im} \epsilon_n >0$,
$\epsilon_n(g) = w \cos(k_n +ig) - i \gamma$. 
It then follows immediately from Eq. (\ref{11}) that the disordered averaged 
inverse complex current 
$J_n = |<J_n^{-1}>^{-1}|$ is
\be
J_n = -i \;  w \;  \sin(k_n +ig)
\ee
in agreement with a direct calculation of the current from the 
spectrum itself. If we focus attention on $ {\rm Im} \; J$ as a function of 
${\rm Re} \; \epsilon$, we obtain 
${\rm Im} J = -w \;  tanh(g)  \;  {\rm Re} \epsilon,$
in excellent qualitative agreement with the results 
obtained for  the imaginary  current with  bounded disorder  in Ref.
\cite{hat} for different values of $g$. Although we have checked our results
 with the Lloyd model, we stress that Eq. (\ref{11}) 
expressing the complex current in terms 
of the Green's function of the Hermitian reference system is  much
more general. 

Fig. (2) shows a typical eigenvalue spectrum for $\L'$ in one dimension  
superimposed  on the winding numbers associated with the extended 
eigenfunctions. The winding numbers are exactly 
$ \pm N/2$ at the band center, and their magnitudes decrease
monotonically as one moves toward the upper edge of the band, i.e., toward 
the lowest energies of the corresponding Hamiltonian. The winding numbers 
remain finite  up to the mobility edge separating complex and real 
eigenvalues, but become undefined in the band tails, where all wavefunctions 
are real. In this case, the winding index classification 
can be replaced by counting the nodes in the localized wavefunction 
\cite{thouless}. The size of the inner excluded region of the projected 
wavefunction shown in the inset shrinks to zero at the mobility edge.  
The winding number at the mobility edge decreases smoothly 
to zero as $g$ increases  and all states become delocalized. 

Given that a well defined topological quantum number 
(which plays the role like that of the momentum) exists for the 
extended states in the one dimensional non-Hermitian localization 
problem, it is interesting to speculate whether similar topological invariants
exist for delocalized states in higher dimensions. For large $g$
 in $D \geq 2$, disorder produces  nontrivial changes in the spectrum, 
but the eigenfunctions are still slightly perturbed plane waves \cite{NS}. 
For  $N \times  N$ lattices with toroidal boundary 
conditions, it seems likely that the winding number $(n_x,n_y)$ embodied 
in the wavevector ${\bf k} = (k_x,k_y) = (2 \pi n_x/N, 2 \pi n_y/N )$
is initially preserved as ${\bf  g}$ decreases from large values. 
However, no detailed analysis exists of how these 
extended states in $D=2$ become localized as their eigenvalues 
approach the real axis, and we cannot be certain if the winding 
number  remains strictly invariant along a trajectory. 
In this context, it is interesting to note 
that a recent analysis of  delocalized wavefunctions 
for the non-Hermitian  problem in $D=2$ predicts algebraic 
decay of the wavefunction amplitude \cite{mudry}, similar to the 
quasi-long range  order below  the vortex 
unbinding transition in statistical 
mechanics. Vortex pairs 
may also  be involved in changes of the vector winding number of delocalized 
wave functions for $d = 2$, similar to the decay of supercurrents in
Helium films
at finite temperature \cite{he}.             
 
It is a pleasure to acknowledge conversations with B. I. Halperin, 
C. Mudry, B. Simons,  J. Avron and   A. Zee. This research was supported by 
the National Science Foundation through Grant No. DMR97-14725 and by 
the Harvard Materials Research Science and Engineering Laboratory 
Through Grant No. DMR94-00396. One of us (N.M.S.) acknowledges  
the  support of Bar-Ilan University.

\end{multicols}{2}

\begin{references}
\bibitem{note} For vortices interacting with columnar defects 
in a cylindrical shell, as in Ref. \cite{hat},
 $\sum_{k=1}^N h_k $ is in fact  entirely determined by
 the perimeter of shell, and there are no fluctuations.
\end{references}
\end{document}